\pdfoutput=1

\documentclass[11pt,twoside,a4paper,cmspaper,final,collab]{cms-tdr}
\def\svnVersion{355fa73-D}\def\svnDate{2019/11/01}\def\cmsCernNoTag{CERN-EP-2019-075}\def\cmsCernDate{\today}\def\cmsMessage{"Published in Physical Review C as \href{http://dx.doi.org/10.1103/PhysRevC.101.014912}{\doi{10.1103/PhysRevC.101.014912}.}"}

\begin{document}\cmsNoteHeader{HIN-17-004}

\hyphenation{had-ron-i-za-tion}
\hyphenation{cal-or-i-me-ter}
\hyphenation{de-vices}
\RCS$HeadURL$
\RCS$Id$

\newlength\cmsFigWidth
\ifthenelse{\boolean{cms@external}}{\setlength\cmsFigWidth{0.5\textwidth}}{\setlength\cmsFigWidth{0.8\textwidth}}
\ifthenelse{\boolean{cms@external}}{\providecommand{\cmsLeft}{upper\xspace}}{\providecommand{\cmsLeft}{left\xspace}}
\ifthenelse{\boolean{cms@external}}{\providecommand{\cmsRight}{lower\xspace}}{\providecommand{\cmsRight}{right\xspace}}

\newcommand {\pp}  {\ensuremath{\text{pp}}\xspace}
\newcommand {\pPb}  {\ensuremath{\text{{\Pp}Pb}}\xspace}
\newcommand {\PbPb}  {\ensuremath{\text{PbPb}}\xspace}
\newcommand {\noff}    {\ensuremath{N_\text{trk}^\text{offline}}\xspace}
\newcommand {\trento}{T$\mathrel{\protect\raisebox{-2.1pt}{R}}$ENTo}
\newcommand {\cn}[1]{c_n\{#1\}}
\newcommand {\vn}[1]{\ensuremath{v_n\{#1\}}}
\newcommand {\dmean}[1]{\langle\langle #1 \rangle\rangle}

\cmsNoteHeader{HIN-17-004}

\title{Multiparticle correlation studies in \texorpdfstring{\pPb}{pPb} collisions at \texorpdfstring{\sqrtsNN = 8.16\TeV}{sqrt(s[NN]) = 8.16 TeV}}

\date{\today}

\abstract{
The second- and third-order azimuthal anisotropy Fourier harmonics of charged particles produced in \pPb collisions, at $\sqrtsNN=8.16\TeV$, are studied over a wide range of event multiplicities. Multiparticle correlations are used to isolate global properties stemming from the collision overlap geometry. The second-order ``elliptic'' harmonic moment is obtained with high precision through four-, six-, and eight-particle correlations and, for the first time, the third-order ``triangular'' harmonic moment is studied using four-particle correlations. A sample of peripheral \PbPb collisions at $\sqrtsNN=5.02\TeV$ that covers a similar range of event multiplicities as the \pPb results is also analyzed. Model calculations of initial-state fluctuations in \pPb and \PbPb collisions can be directly compared to the high precision experimental results. This work provides new insight into the fluctuation-driven origin of the $v_3$ coefficients in \pPb and \PbPb collisions, and into the dominating overall collision geometry in \PbPb collisions at the earliest stages of heavy ion interactions.
}

\hypersetup{%
pdfauthor={CMS Collaboration},%
pdftitle={Multiparticle correlation studies in pPb collisions at sqrt(s[NN]) = 8.16 TeV},%
pdfsubject={CMS},%
pdfkeywords={CMS, heavy ion, cumulant, flow, collectivity, small system}}

\maketitle

\section{Introduction}

In collisions of ultra relativistic heavy ions, two-particle azimuthal correlations between the large number of particles created over a broad range in pseudorapidity, were first observed in gold--gold and
copper--copper collisions at the BNL RHIC~\cite{Alver:2008aa,Adams:2005ph,Abelev:2009af,Alver:2009id}, and have subsequently been studied in lead--lead (\PbPb) collisions at the CERN LHC~\cite{Chatrchyan:2011eka,ALICE:2011ab,Aamodt:2011by,Chatrchyan:2012wg,Aamodt:2010pa,ATLAS:2012at,Chatrchyan:2012ta}.
These correlations are  thought to reflect the collective motion of a strongly interacting and expanding medium with quark and gluon degrees of freedom, namely, the quark-gluon plasma~\cite{Busza:2018rrf}.
The observed azimuthal correlation structure can be characterized by Fourier harmonics, with the second ($v_2$) and third ($v_3$)  harmonics referred to as ``elliptic'' and ``triangular'' flow, respectively.
Within a hydrodynamic picture, the Fourier harmonics are related to the initial geometry of the colliding system and provide insight into the transport properties of the produced medium~\cite{Alver:2010dn,Schenke:2010rr,Qiu:2011hf}.
Fluctuations can also arise from the discrete substructure of the interaction region at the parton level~\cite{Alver:2010gr,Miller:2007ri} and can have a significant effect on the observed higher-order harmonic coefficients.

Two-particle azimuthal correlations, which are long-range in pseudorapidity, are also found in small systems for collisions leading to  high final-state particle densities.
At the LHC, long-range correlations have been observed in proton--proton (\pp)~\cite{Khachatryan:2010gv,Aad:2015gqa,Khachatryan:2015lva} and proton--lead (\pPb)~\cite{CMS:2012qk,Abelev:2012ola,Aad:2012gla,Aaij:2015qcq} collisions.
Similar results have been obtained  at RHIC in studies of deuteron--gold, proton--gold, and helium-3--gold collisions~\cite{Aidala:2017ajz,Aidala:2016vgl,Adare:2015ctn,PHENIX:2018lia}.
The origin of the long-range correlations in  systems involving only a small number of participating nucleons is still under active discussion~\cite{Nagle:2018nvi}.
One possibility is that fluctuation-driven asymmetries in the initial-state nucleon locations within the overlap region are transferred to the final-state particle distributions through the hydrodynamic evolution of an expanding plasma~\cite{Schenke:2014zha,Bozek:2011if,Bozek:2012gr}.  Alternatively, it has been proposed that the observed behavior arises from
the transfer of initial-state gluon correlations to the produced particles~\cite{Dusling:2012wy,Dusling:2017dqg,Dusling:2017aot}.

Studies of azimuthal correlations in small systems using two or more particles, as achieved through the use of a multiparticle cumulant expansion~\cite{Borghini:2000sa},  show that the \pp~\cite{Khachatryan:2016txc} and \pPb~\cite{Khachatryan:2015waa} systems develop  similar  collective behavior to that found in heavier systems~\cite{Chatrchyan:2013kba}.
The cumulants quantify an $n$th order contribution of the azimuthal correlation that is irreducible to lower-order correlations.
By requiring correlations among multiple particles, correlations that are not related to a bulk property of the medium, such as back-to-back jet correlations and resonance decays, are strongly suppressed~\cite{Chatrchyan:2013nka}.
The $v_n$ harmonics based on different orders of the multiparticle expansion provide information on the event-by-event fluctuation of the observed anisotropy~\cite{Bilandzic:2010jr}.
Previous $v_2$ multiparticle cumulant results for \pPb collisions at a nucleon-nucleon center-of-mass energy of $\sqrtsNN = 5.02\TeV$ suggest a
direct correlation of the final-state asymmetry with the initial-state eccentricity of the participating nucleons~\cite{Khachatryan:2015waa,Yan:2013laa}.
The $v_3$ harmonic is expected to be dominated by fluctuations in the initial-state geometry.
The multiparticle correlations of the $v_3$ harmonic are then expected to reflect these fluctuations.
An earlier multiparticle correlation measurement by the ATLAS Collaboration has found evidence for a finite $v_3$ harmonic amplitude in the \pPb system~\cite{Aaboud:2017blb}.
With precise measurements of the $v_2$ and $v_3$ multiparticle cumulants, it becomes possible to make direct comparison of calculations based on eccentricity fluctuations in the initial-state geometry to the higher-order moments of the fluctuation distributions.
The measurements provide key input for models that explore the hydrodynamic expansion of the medium~\cite{Heinz:2013th,Giacalone:2017uqx}, as well as for models that propose that final-state asymmetries in light systems arise from partons scattering off localized domains of color charge in the initial state~\cite{Dusling:2017dqg}.
In the hydrodynamic picture, the $v_2$ and $v_3$ values are dominated by fluctuations in \pPb collisions.
In \PbPb collisions, the $v_2$ value is dominated by the lenticular shape of the overlap geometry, while the $v_3$ value is dominated by initial-state fluctuations of the nucleon locations~\cite{Alver:2010gr}.

In this work, the results from \pPb collisions at $\sqrtsNN = 8.16\TeV$ are studied with a significant improvement in the  precision of the $v_{2}$ results compared to the earlier measurements at  $\sqrtsNN = 5.02\TeV$. For the first time,  the $v_3$ harmonic is determined by multiparticle correlations.
The \pPb results are also compared to those found for \PbPb collisions at $\sqrtsNN = 5.02\TeV$ to explore the dependence on the overlap geometry.
The ratios between the four-particle and two-particle $v_n$ values provide information on the relative importance of the global geometry and the fluctuation-driven asymmetries~\cite{Ollitrault:2009ie}.
These ratios are explored for both the $v_2$ and $v_3$ harmonics and are compared between the \pPb and \PbPb systems.

\section{Experimental setup and data sample}

The CMS detector comprises a number of subsystems~\cite{JINST}.
The results in this paper are mainly based on the
silicon tracker information. The silicon tracker, located in the 3.8\unit{T} field of a
superconducting solenoid, consists of 1\,440 silicon pixel and 15\,148 silicon strip
detector modules. The silicon tracker measures charged particles within the laboratory pseudorapidity
range $\abs{\eta}< 2.5$, and provides an impact parameter resolution of ${\approx}15\mum$ and
a transverse momentum (\pt) resolution better than 1.5\% up to 100\GeVc~\cite{JINST}.
The electromagnetic (ECAL) and hadron (HCAL) calorimeters are also
located inside the solenoid and cover the pseudorapidity range $\abs{\eta} < 3.0$.
The HCAL barrel and endcaps are sampling calorimeters composed of brass and
scintillator plates.
The ECAL consists of lead tungstate crystals arranged in a quasi-projective geometry.
Iron and quartz-fiber \v Cerenkov hadron forward (HF) calorimeters cover the range $3.0 < \abs{\eta} < 5.2$ on either
side of the interaction region.
These HF calorimeters are azimuthally subdivided into $20^{\circ}$
modular wedges and further segmented to form $0.175{\times}0.175$ rad $(\Delta\eta{\times}\Delta\phi)$ cells.
The ECAL and HCAL cells are grouped to form ``towers.''
The detailed Monte Carlo (MC) simulation of the CMS detector response is based
on \GEANTfour~\cite{GEANT4}.

The analysis is performed using data recorded by CMS during the LHC \pPb run in 2016 and corresponds to an integrated luminosity of 186\nbinv~\cite{CMS-PAS-LUM-17-002}.
The beam energies were 6.5\TeV for protons and 2.56\TeV per nucleon for lead nuclei, resulting in $\sqrtsNN = 8.16\TeV$.
The beam directions were reversed during the run allowing a check of potential detector related systematic uncertainties. No significant differences were detected and the merged results are reported.
The nucleon-nucleon center-of-mass in the \pPb collisions is not at rest with respect to the laboratory frame because of the energy difference between the colliding particles.
Massless particles emitted at $\eta_\text{cm} = 0$ in the nucleon-nucleon center-of-mass frame will be detected at $\eta = -0.465$ (clockwise proton beam)  or $0.465$ (counterclockwise proton beam) in the laboratory frame.
In this paper, an unsubscripted $\eta$ symbol is used to denote the laboratory frame pseudorapidity.
A sample of $\sqrtsNN = 5.02\TeV$ \PbPb data collected during the 2015 LHC heavy ion run, corresponding to an integrated luminosity of 1.2\mubinv, is also analyzed for
comparison purposes. The triggers and event selection, as well as track reconstruction
and selection, are identical to those used in Ref.~\cite{Sirunyan:2017uyl} and are summarized below.

Minimum bias (MB) \pPb events were triggered by requiring at least
one track with $\pt > 0.4\GeVc$ in the pixel tracker
during a \pPb bunch crossing and the presence of at least one tower in one of the two HF detectors having an energy above 1\GeV.
In order to select high-multiplicity \pPb collisions,
a dedicated high-multiplicity trigger was implemented using the CMS level-1
(L1) and high-level trigger (HLT) systems~\cite{Khachatryan:2016bia}.
At L1, the total number of ECAL and HCAL towers with the transverse energies above a threshold of 0.5\GeV is required to exceed 120 and 150 in ECAL and HCAL, respectively.
The events which pass the L1 trigger are then subsequently filtered in the HLT.
The track reconstruction that is performed online, as part of the HLT trigger, uses the identical reconstruction algorithm as employed in the offline processing~\cite{Chatrchyan:2014fea}.
For each event, the vertex reconstructed with the highest number of pixel detector tracks was selected.
The number (multiplicity) of pixel tracks (${N}_\text{trk}^\text{online}$)
with $\abs{\eta}<2.4$, $\pt > 0.4\GeVc$, and a distance of closest approach to this vertex of 0.4\cm or
less, was determined for each event. Several multiplicity ranges were defined with
prescale factors that were reduced with increasing particle multiplicity until, for the highest-multiplicity events, no prescale was applied.

In the offline analysis, hadronic collisions are selected by the requirement of a
coincidence of at least one HF calorimeter tower containing more than 3\GeV of
total energy in each of the HF detectors within $3.0<\abs{\eta}<5.2$.
Events are also required to contain at least one reconstructed primary
vertex within 15\cm from the nominal interaction point along the beam axis
and within 0.15\cm transverse to the beam trajectory.
At least two reconstructed tracks are required to be associated
with the primary vertex. Beam-related background is suppressed
by rejecting events for which less than 25\% of all reconstructed tracks
pass the track selection criteria.

Tracks are used that pass the high-purity selection criteria described in Ref.~\cite{Chatrchyan:2014fea}.
In addition, a reconstructed track is only considered as a
candidate track from the primary vertex if the separation along the beam axis ($z$)
between the track and the best vertex, and
the track-vertex impact parameter measured transverse to the beam
are each less than three times their respective uncertainties. The relative uncertainty
in the \pt measurement is required to be less than 10\%.
To restrict the analysis to a kinematic region of high tracking efficiency and a low rate of incorrectly reconstructed tracks,
only tracks with $\abs{\eta}<2.4$ and  $0.3 < \pt < 3.0\GeVc$ are used.

The entire \pPb data set is divided into classes of reconstructed track multiplicity, \noff,
where primary tracks passing the high-purity criteria and with $\abs{\eta}<2.4$ and $\pt >0.4\GeVc$ are counted.
The HLT \pt cutoff, which is only applied on determination of \noff, is higher than that used for the analysis because of online processing time constraints.
The absence of the time constraints in the offline process allows extending the \pt coverage down to 0.3\GeVc in the cumulant calculation.
The multiplicity classification in this analysis is identical to that used in Ref.~\cite{Chatrchyan:2013nka},
where more details are provided, including a table relating \noff\ to the fraction of MB triggered events.
The \PbPb sample is reprocessed using the same event selection and track reconstruction
as for the present \pPb analysis.
A description of the analysis of 2015 \PbPb data can be found in Ref.~\cite{Sirunyan:2017uyl}.

\section{Analysis techniques}

The analysis is done using the $Q$-cumulant method~\cite{Bilandzic:2010jr}.
Here it is possible to determine the $n$th harmonic moment based on correlations among all possible grouping of $m$ particles, where $m$ also corresponds to the cumulant order.
The multiparticle correlations for cumulant orders 2 through 8 can be expressed as:
\begin{equation*}
\begin{aligned}
\dmean{2} &\equiv
    \left<\!\left< \re^{in(\phi_{1} - \phi_{2}} \right>\!\right>,
\\
\dmean{4} &\equiv
    \left<\!\left< \re^{in(\phi_{1} + \phi_{2} - \phi_{3} - \phi_{4}} \right>\!\right>,
\\
\dmean{6} &\equiv
    \left<\!\left< \re^{in(\phi_{1} + \phi_{2} + \phi_{3} - \phi_{4} - \phi_{5}
    - \phi_{6})} \right>\!\right>,
\\
\dmean{8} &\equiv
    \left<\!\left< \re^{in(\phi_{1} + \phi_{2} + \phi_{3} + \phi_{4} - \phi_{5} - \phi_{6} - \phi_{7}
    - \phi_{8})} \right>\!\right>,
\end{aligned} \label{eq:corr}
\end{equation*}
where $\phi_{i}$ $(i=1,\ldots,m)$ are the
azimuthal angles of one unique combination of $m$ particles in an event, $n$ is the harmonic number (2 for elliptic and 3 for triangular flow, respectively),
and $\left\langle\left\langle \cdots \right\rangle\right\rangle$ represents
the average over all combinations from all events within a given \noff range.
The higher-order cumulants, $\cn{m}$, are calculated as~\cite{Bilandzic:2010jr}
\begin{equation*}
\begin{aligned}
\cn{4} &= \dmean{4} - 2 \dmean{2}^2,
\\
\cn{6} &= \dmean{6} - 9 \dmean{4}\dmean{2} + 12 \dmean{2}^3,
\\
\cn{8} &= \dmean{8} - 16 \dmean{6}\dmean{2} -18 \dmean{4}^2
\\
       &+ 144 \dmean{4}\dmean{2}^2 - 144 \dmean{2}^4.
\end{aligned} \label{eq:cn}
\end{equation*}
The Fourier harmonics $v_n\{m\}$ that characterize the global azimuthal behavior
can be related to the $m$-particle correlations using a generic framework discussed in Ref.~\cite{Bilandzic:2013kga}, with
\begin{equation*}
\begin{aligned}
\vn{4} &= \sqrt[4]{-\cn{4}},
\vn{6} &= \sqrt[6]{\frac{1}{4} \cn{6}},\\
\vn{8} &= \sqrt[8]{-\frac{1}{33} \cn{8}}.
\end{aligned}\label{eq:vn}
\end{equation*}
Each reconstructed track is weighted by a correction factor to account
for the reconstruction efficiency, the detector acceptance, and
the fraction of misreconstructed tracks. This factor is based on \HIJING 1.383~\cite{Gyulassy:1994ew} MC simulations, and is determined  as a
function of \pt, $\eta$, and $\phi$, as described in Refs.~\cite{Chatrchyan:2011eka,Chatrchyan:2012wg}.
The $\cn{4}$ and $\cn{8}$ values need to be negative, and the $\cn{6}$ value needs to be positive, in order to have real values for the $\vn{m}$ coefficients.
The same method has been used in previous CMS analyses~\cite{Chatrchyan:2013nka,Khachatryan:2015waa,Sirunyan:2017igb}.
The two-particle correlation $v_n\{2\}$ can be measured as described in Ref.~\cite{Sirunyan:2017uyl}.
Increasing the numbers of particles used to determine the correlations for a given harmonic reduces the sensitivity of the results to few-particle correlations that are not related to a global behavior.
The ratios between $v_n$ harmonics involving different number of particles can be used to test the system independence of fluctuation-driven initial-state anisotropies in the hydrodynamic picture.
In particular, the triangular flow ratio $v_3\{4\}/v_3\{2\}$, which is dominated by fluctuations, can be used to confirm this expectation.

A number of potential sources of systematic uncertainties affecting the experimental $v_n\{m\}$ values are considered.
The sensitivity of the results to the selection criteria for valid tracks was studied by varying the criteria. The sensitivity to the primary vertex position was explored by performing the analysis for different vertex $z$ ranges.
The potential for an HLT trigger bias was investigated by changing the  trigger thresholds.  Pileup effects, where two or more interactions occur in the same bunch crossing, were studied by comparing results obtained during different beam differential luminosity periods.  For the \pPb results, the beam directions were reversed, allowing for potential detector acceptance effects to be explored.
No evident \noff\ dependent systematic effects are observed.
The total systematic uncertainties, obtained by combining the individual uncertainties in quadrature, are found to be 1--2.4\% for the $v_2$ coefficients for both \pPb and \PbPb collisions and  5 (2.6)\% for the \pPb (\PbPb) $v_3$ results.
The \pPb (\PbPb) $v_{n}\{4\}/v_{n}\{2\}$ ratios systematic uncertainties are found to be 3 (1)\%.
The \pPb $v_2\{6\}/v_2\{4\}$ and $v_2\{8\}/v_2\{6\}$ ratios systematic uncertainties are found to be 3\%.

\section{Results}

\begin{figure}[htb]
	\centering
	\includegraphics[width=\cmsFigWidth]{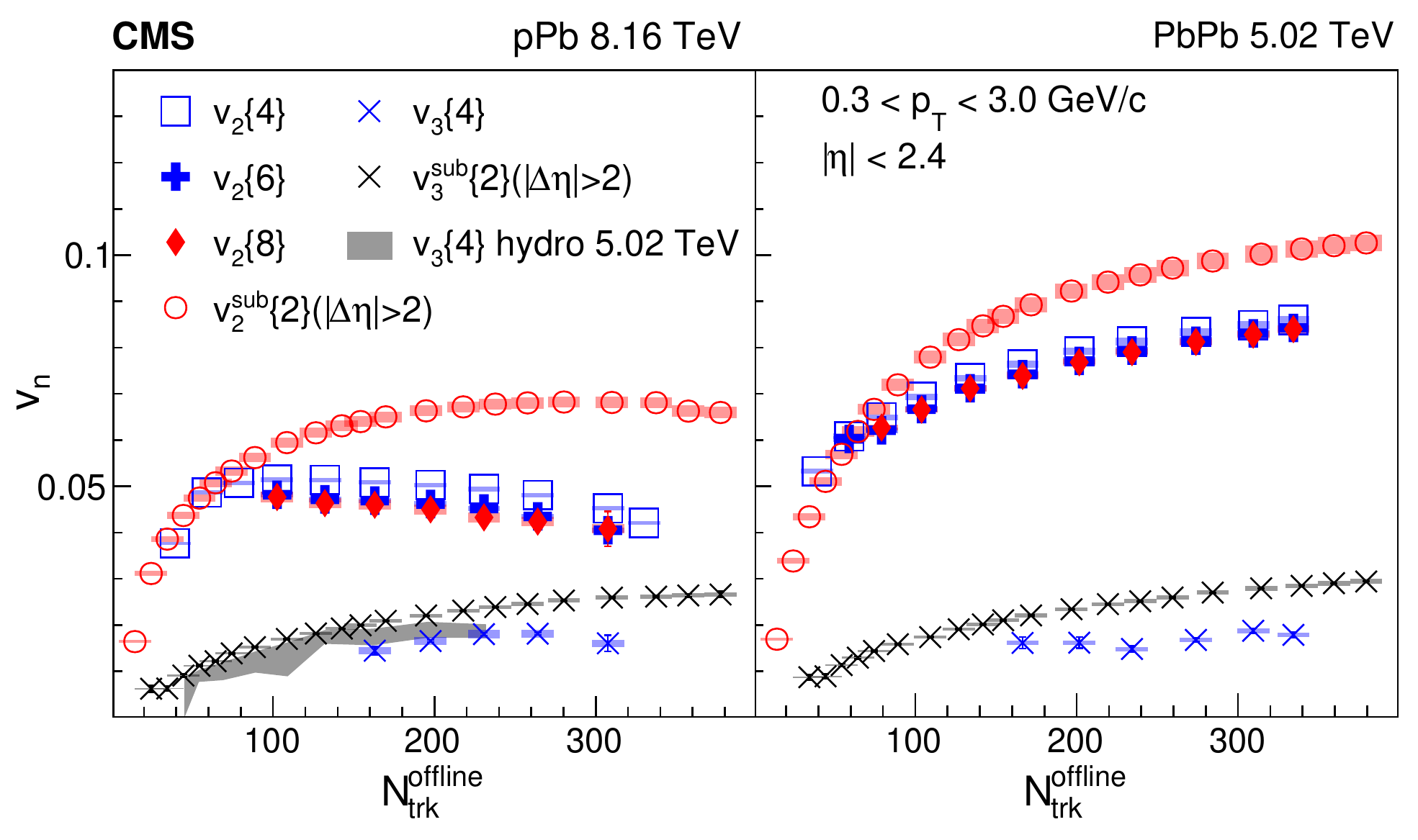}
	\caption{The multiparticle $v_2\{4,6,8\}$ and $v_3\{4\}$ harmonics are shown for \pPb 8.16\TeV (left) and \PbPb 5.02\TeV (right) collisions as a function of \noff.
		Two-particle results $v_2^{\text{sub}}\{2\}(\abs{\Delta\eta}>2)$ and $v_3^{\text{sub}}\{2\}(\abs{\Delta\eta}>2)$ are from Ref.~\cite{Sirunyan:2017uyl}.
		Error bars and shaded boxes denote statistical and systematic uncertainties, respectively.
		The shaded area shows the hydrodynamic prediction of $v_3\{4\}$ in \pPb collisions at $\sqrtsNN = 5.02\TeV$~\cite{Kozlov:2014fqa}.
		}
	\label{fig:vn}
\end{figure}

The second- and third-order harmonic multiparticle cumulant results $v_2$ and $v_3$ for charged particles with $0.3 < \pt < 3.0\GeVc$ and $\abs{\eta}<2.4$ are shown in Fig.~\ref{fig:vn} for \pPb collisions at $\sqrtsNN = 8.16\TeV$ and for \PbPb collisions at $\sqrtsNN = 5.02\TeV$.
The two-particle correlation results $v_2^{\text{sub}}\{2\}(\abs{\Delta\eta}>2)$ and $v_3^{\text{sub}}\{2\}(\abs{\Delta\eta}>2)$, with low-multiplicity subtraction to remove jet correlations, are taken from Ref.~\cite{Sirunyan:2017uyl}.
The multiparticle elliptic flow harmonics $v_2\{4\}$, $v_2\{6\}$, and $v_2\{8\}$ are found to be real and of similar magnitude.
The four-particle triangular flow harmonic, $v_3\{4\}$, is also found to be real, with an amplitude consistent with the earlier ATLAS $c_3\{4\}$ results~\cite{Aaboud:2017blb}.
These results indicate collective behavior in high multiplicity \pPb collisions at $\sqrtsNN = 8.16\TeV$~\cite{Heinz:2013th,Giacalone:2017uqx}.
Comparing the different systems, the $v_2$ values for \PbPb collisions are higher than those for \pPb collisions, which is consistent with the lenticular-shaped overlap geometry dominating this harmonic for \PbPb collisions.
The two-particle correlation $v_2$ and $v_3$ results are systematically higher than the multiparticle results for both \pPb and \PbPb collision.
This is expected if there is a significant fluctuation component,
which is expected to increase the two-particle correlation results and decrease the multiparticle correlation results, as compared to case where fluctuations are absent~\cite{Ollitrault:2009ie}.
With increasing \noff, the $v_2\{4\}$, $v_2\{6\}$, and $v_2\{8\}$ values all rise in \PbPb collisions, while they fall slightly in \pPb collisions.
This might suggest that the fluctuation-driven component of the eccentricity, as compared to the component arising from the lenticular overlap geometry, is decreasing with increasing multiplicity in the \PbPb system.
The $v_3$ values are comparable for both systems, as expected if this higher-order harmonic is dominated by fluctuation behavior.
A (3+1)D event-by-event viscous hydrodynamic calculation of the four-particle cumulant $v_3\{4\}$ for \pPb collisions at $\sqrtsNN = 5.02\TeV$~\cite{Kozlov:2014fqa} is also shown in Fig.~\ref{fig:vn} as a gray band.
This calculation, with an entropy distribution taken as a two-dimensional Gaussian of width $\sigma = 0.4$ fm and having a shear viscosity-to-entropy ratio of $\eta/s = 0.08$, is found to be consistent with the data.

\begin{figure}[htb]
	\centering
	\includegraphics[width=\cmsFigWidth]{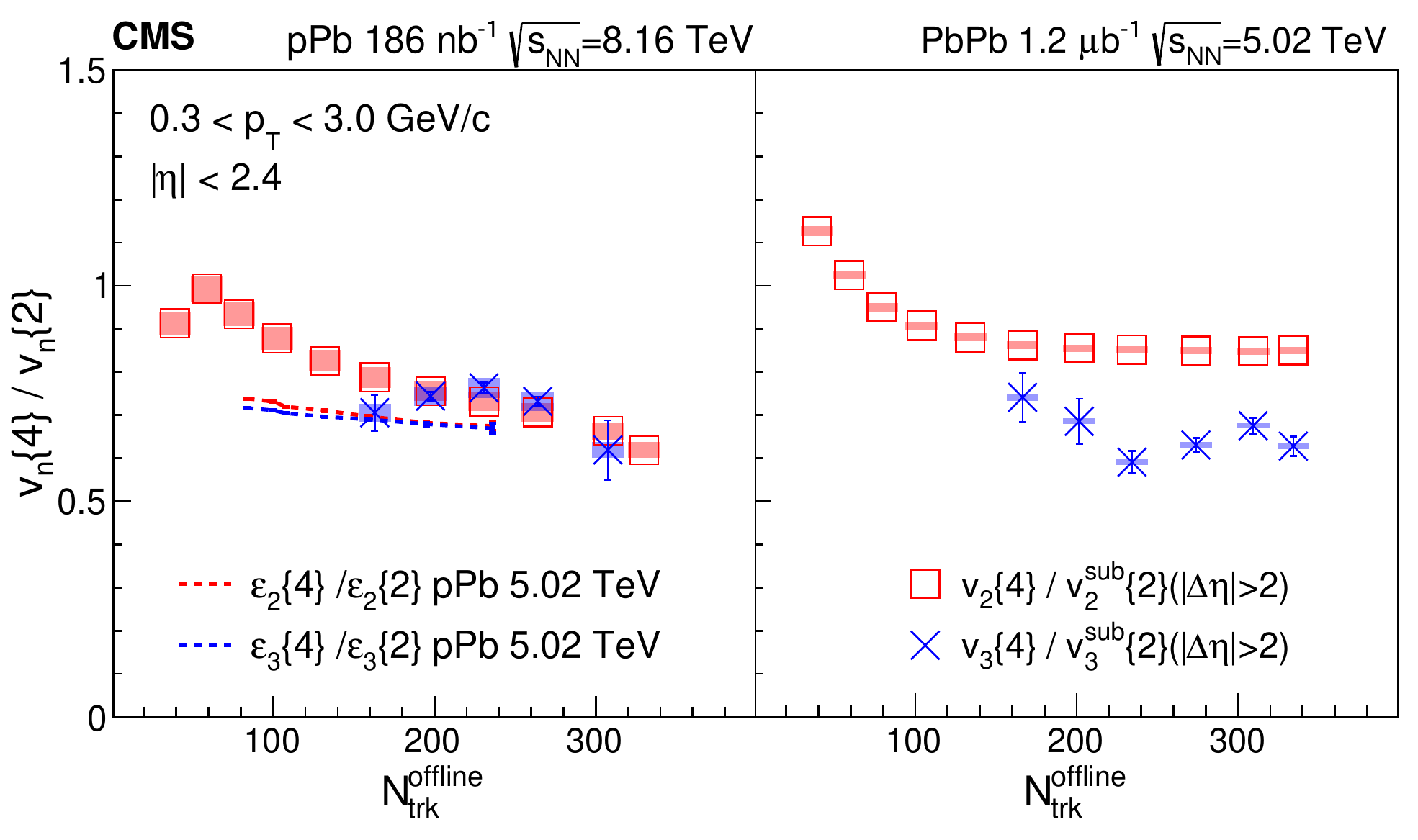}
	\caption{The ratios of four- and two-particle harmonics ($v_2\{4\}/v_2\{2\}$ and $v_3\{4\}/v_3\{2\}$) are shown for \pPb $\sqrtsNN = 8.16\TeV$ (left) and \PbPb $\sqrtsNN = 5.02\TeV$ (right) collisions as a function of \noff.
		Error bars and shaded boxes denote statistical and systematic uncertainties, respectively.
		The dashed curves show a hydrodynamics-motivated initial-state fluctuation calculation of eccentricities $\varepsilon_{n}\{m\}$ for \pPb collisions at $\sqrtsNN = 5.02\TeV$~\cite{Giacalone:2017uqx}.
		}
	\label{fig:v42}
\end{figure}

Figure~\ref{fig:v42} shows the ratios $v_2\{4\}/v_2\{2\}$ and  $v_3\{4\}/v_3\{2\}$  for both the \pPb and \PbPb systems.
For \pPb collisions, the ratios for $v_2$ and $v_3$ are similar within uncertainties, which is consistent with having both the second- and third-order harmonics arising from the same initial-state fluctuation mechanism.
Comparing the \pPb and \PbPb systems,  the $v_3$ ratios are comparable for both systems, while the $v_2$ ratios are higher in \PbPb than in \pPb for higher \noff values, again reflecting the larger geometric contribution for the heavier system collisions.
The $v_2$ ratio for \PbPb collisions saturates at large multiplicity while, for \pPb collisions, the ratio continues to decrease as the multiplicity increases.

Cumulants can also be constructed for the eccentricities of the matter distribution in the initial state, $\varepsilon_{n}\{m\}$.
In the hydrodynamic picture, the $v_n\{m\}$ values are proportional to $\varepsilon_{n}\{m\}$, with $v_{n}\{m\}=k_{n}\varepsilon_{n}\{m\}$, where $k_{n}$ reflects the medium properties and does not depend on the order of the cumulant. Therefore, ratios of different cumulant $v_n$ values can directly probe properties of initial-state eccentricity.
This is shown in Fig.~\ref{fig:v42} based on Glauber model initial condition simulated using the \trento\ framework~\cite{Bernhard:2016tnd}, and assuming a width $\sigma=0.3$\unit{fm} of the source associated with each nucleon~\cite{Giacalone:2017uqx}.
The calculations were done for \pPb collisions at $\sqrtsNN = 5.02\TeV$ by varying the geometric overlap of the colliding nuclei.
It should be noted that the two-particle correlation results were obtained with a large pseudorapidity gap of $\abs{\Delta\eta}>2$.
Earlier experimental \pPb results at $\sqrtsNN = 5.02\TeV$ have shown that this gap can lead to a reduction in the observed $v_2^{\text{sub}}\{2\}(\abs{\Delta\eta}>2)$ values by 10\% resulting from event-plane fluctuations~\cite{Khachatryan:2015oea}.
This gap dependence is not directly determined in the current measurement and, consequently, the reported values are not corrected for this effect.
However, assuming a 10\% gap-related reduction in the two-particle $v_2\{2\}$ values with, in the absence of a gap, the $v_2\{4\}$ values not being similarly affected, the reported values of $v_2\{4\}/v_2\{2\}$ might be too high by 10\%.

\begin{figure}[htb]
	\centering
	\includegraphics[width=0.7\cmsFigWidth]{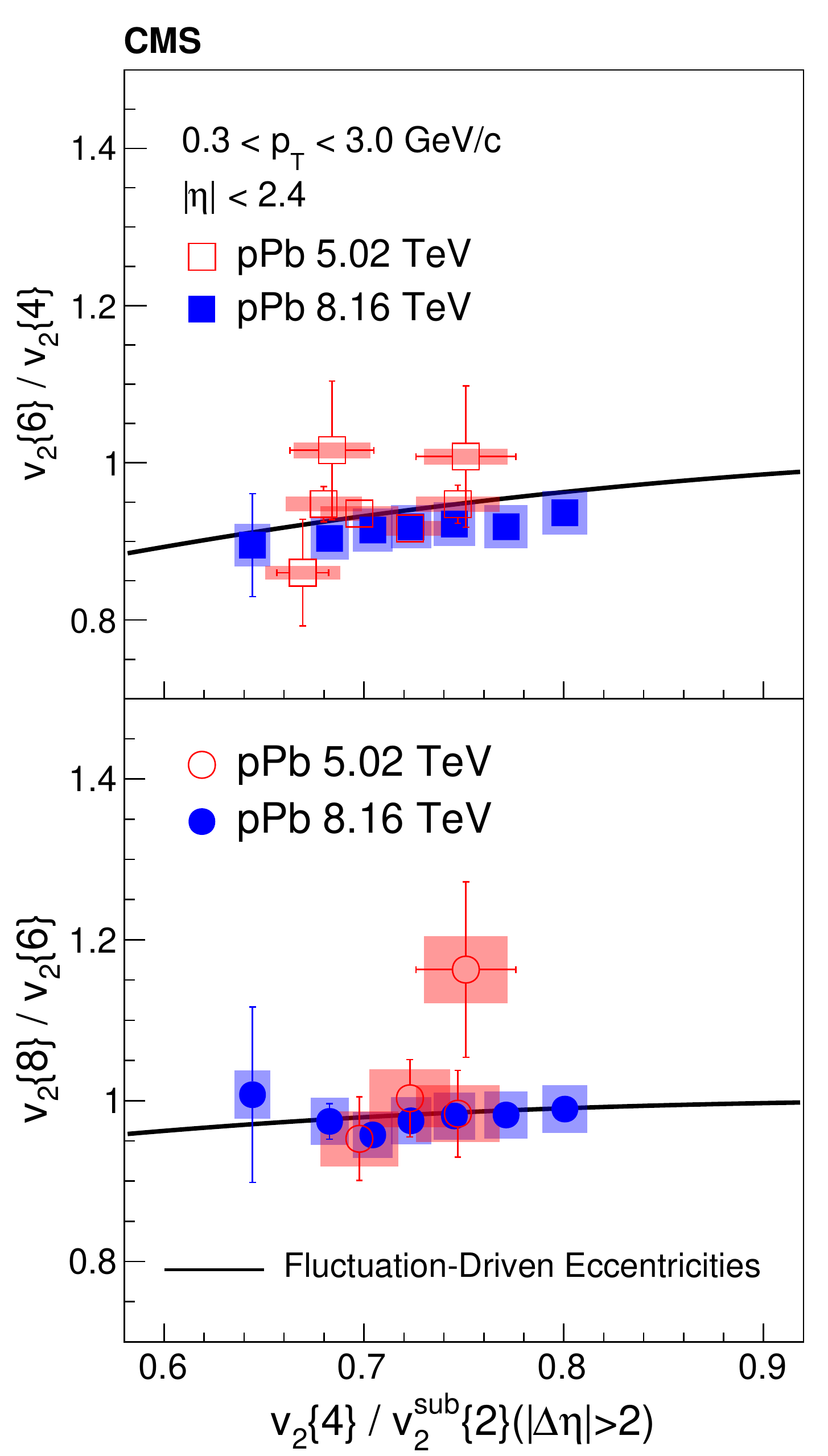}
	\caption{ Cumulant ratios $v_2\{6\}/v_2\{4\}$ (upper) and $v_2\{8\}/v_2\{6\}$ (lower) as a function of $v_2\{4\}/v^{\text{sub}}_2\{2\}$ in \pPb collisions at $\sqrtsNN = 5.02\TeV$~\cite{Khachatryan:2015waa} and 8.16\TeV.
	Error bars and shaded areas denote statistical and systematic uncertainties, respectively.
	The solid curves show the expected behavior based on a hydrodynamics-motivated study of the role of initial-state fluctuations~\cite{Yan:2013laa}.
	}
	\label{fig:v2rs}
\end{figure}

In Fig.~\ref{fig:v2rs}, the ratios $v_{2}\{6\}/v_2\{4\}$ and $v_{2}\{8\}/v_2\{6\}$ are shown as a function of the ratio $v_{2}\{4\}/v_2\{2\}$ for \pPb collisions at $\sqrtsNN = 8.16\TeV$ and compared to calculations based on fluctuation-driven eccentricities~\cite{Yan:2013laa} with a universal power law distribution assumed for the eccentricities instead of a two-dimensional Gaussian distribution.
These results are similar to those previously reported in Ref.~\cite{Khachatryan:2015waa} for \pPb at $\sqrtsNN = 5.02\TeV$, as shown in the figure, but with greatly reduced statistical uncertainties.
Within the uncertainties, the model calculations for both the  $v_2\{6\}/v_2\{4\}$ and $v_2\{8\}/v_2\{6\}$ ratios agree with the experimental results.
The agreement improves if the reduced correlation resulting from the $v_2^{\text{sub}}\{2\}(\abs{\Delta\eta}>2)$ pseudorapidity gap is also considered.
The agreement of the calculations with the data shows that the differences found among the multiparticle cumulant results for the $v_2$ harmonic can be described by non-Gaussian initial-state fluctuations.
The precise measurement of the ratio results confirms the hypothesis that the multiparticle correlations originate from the product of single-particle correlations arising from source fluctuations with respect to overall collision geometry.
This is a fundamental assumption of both the hydrodynamic~\cite{Giacalone:2017uqx} and the Color Glass Condensate model calculations~\cite{Dusling:2017dqg}.

\section{Summary}

In summary, the azimuthal anisotropy for \pPb collisions at $\sqrtsNN = 8.16\TeV$ and \PbPb collisions at $\sqrtsNN = 5.02\TeV$ are studied as a function of the final-state particle multiplicities with the CMS experiment.
The $v_2$ Fourier coefficient is determined using cumulants obtained with four-, six-, and eight-particle correlations with greatly increased precision compared to previous measurements.
The higher-order $v_3\{4\}$ coefficient is reported for the first time for a small system.
For  \pPb collisions, the ratios $v_2\{4\}/v_2\{2\}$ and $v_3\{4\}/v_3\{2\}$ are comparable, consistent with a purely fluctuation-driven origin for the azimuthal asymmetry.
Both the \pPb and \PbPb systems have very similar $v_3$ coefficients for the cumulant orders studied, indicating a similar, fluctuation-driven initial-state geometry.
In contrast, both the magnitude of the $v_2$ coefficients and the $v_2\{4\}/v_2\{2\}$ ratio is larger for \PbPb collisions, as expected if the overall collision geometry dominates.
The $v_2$ cumulant ratios for \pPb collisions are consistent with a collective flow behavior  that originates from and is proportional to the initial-state anisotropy.

\begin{acknowledgments}
We congratulate our colleagues in the CERN accelerator departments for the excellent performance of the LHC and thank the technical and administrative staffs at CERN and at other CMS institutes for their contributions to the success of the CMS effort. In addition, we gratefully acknowledge the computing centres and personnel of the Worldwide LHC Computing Grid for delivering so effectively the computing infrastructure essential to our analyses. Finally, we acknowledge the enduring support for the construction and operation of the LHC and the CMS detector provided by the following funding agencies: BMBWF and FWF (Austria); FNRS and FWO (Belgium); CNPq, CAPES, FAPERJ, FAPERGS, and FAPESP (Brazil); MES (Bulgaria); CERN; CAS, MoST, and NSFC (China); COLCIENCIAS (Colombia); MSES and CSF (Croatia); RPF (Cyprus); SENESCYT (Ecuador); MoER, ERC IUT, and ERDF (Estonia); Academy of Finland, MEC, and HIP (Finland); CEA and CNRS/IN2P3 (France); BMBF, DFG, and HGF (Germany); GSRT (Greece); NKFIA (Hungary); DAE and DST (India); IPM (Iran); SFI (Ireland); INFN (Italy); MSIP and NRF (Republic of Korea); MES (Latvia); LAS (Lithuania); MOE and UM (Malaysia); BUAP, CINVESTAV, CONACYT, LNS, SEP, and UASLP-FAI (Mexico); MOS (Montenegro); MBIE (New Zealand); PAEC (Pakistan); MSHE and NSC (Poland); FCT (Portugal); JINR (Dubna); MON, RosAtom, RAS, RFBR, and NRC KI (Russia); MESTD (Serbia); SEIDI, CPAN, PCTI, and FEDER (Spain); MOSTR (Sri Lanka); Swiss Funding Agencies (Switzerland); MST (Taipei); ThEPCenter, IPST, STAR, and NSTDA (Thailand); TUBITAK and TAEK (Turkey); NASU and SFFR (Ukraine); STFC (United Kingdom); DOE and NSF (USA).

\hyphenation{Rachada-pisek} Individuals have received support from the Marie-Curie program and the European Research Council and Horizon 2020 Grant, contract Nos.\ 675440 and 765710 (European Union); the Leventis Foundation; the A.P.\ Sloan Foundation; the Alexander von Humboldt Foundation; the Belgian Federal Science Policy Office; the Fonds pour la Formation \`a la Recherche dans l'Industrie et dans l'Agriculture (FRIA-Belgium); the Agentschap voor Innovatie door Wetenschap en Technologie (IWT-Belgium); the F.R.S.-FNRS and FWO (Belgium) under the ``Excellence of Science -- EOS" -- be.h project n.\ 30820817; the Beijing Municipal Science \& Technology Commission, No. Z181100004218003; the Ministry of Education, Youth and Sports (MEYS) of the Czech Republic; the Lend\"ulet (``Momentum") Program and the J\'anos Bolyai Research Scholarship of the Hungarian Academy of Sciences, the New National Excellence Program \'UNKP, the NKFIA research grants 123842, 123959, 124845, 124850, 125105, 128713, 128786, and 129058 (Hungary); the Council of Science and Industrial Research, India; the HOMING PLUS program of the Foundation for Polish Science, cofinanced from European Union, Regional Development Fund, the Mobility Plus program of the Ministry of Science and Higher Education, the National Science Center (Poland), contracts Harmonia 2014/14/M/ST2/00428, Opus 2014/13/B/ST2/02543, 2014/15/B/ST2/03998, and 2015/19/B/ST2/02861, Sonata-bis 2012/07/E/ST2/01406; the National Priorities Research Program by Qatar National Research Fund; the Programa Estatal de Fomento de la Investigaci{\'o}n Cient{\'i}fica y T{\'e}cnica de Excelencia Mar\'{\i}a de Maeztu, grant MDM-2015-0509 and the Programa Severo Ochoa del Principado de Asturias; the Thalis and Aristeia programs cofinanced by EU-ESF and the Greek NSRF; the Rachadapisek Sompot Fund for Postdoctoral Fellowship, Chulalongkorn University and the Chulalongkorn Academic into Its 2nd Century Project Advancement Project (Thailand); the Welch Foundation, contract C-1845; and the Weston Havens Foundation (USA).
\end{acknowledgments}

\bibliography{auto_generated}

\providecommand{\href}[2]{#2}\begingroup\raggedright\begin{thebibliography}{10}%
\makeatletter
\providecommand{\hrefCMSnoop }[0]{\@secondoftwo}%
\makeatother
\providecommand{\doi}{\texttt{doi:}\begingroup \urlstyle{tt}\Url}

\bibitem{Alver:2008aa}
\hrefCMSnoop {}{{PHOBOS} Collaboration, ``{System size dependence of cluster
  properties from two-particle angular correlations in Cu+Cu and Au+Au
  collisions at $\sqrt{s_{NN}}$ = 200 GeV}'',} \textit{ Phys. Rev. C} \textbf{
  81} (2010) 024904,
  \href{http://dx.doi.org/10.1103/PhysRevC.81.024904}{\doi{10.1103/PhysRevC.81.024904}},
\href{http://www.arXiv.org/abs/0812.1172}{\texttt{arXiv:0812.1172}}.
%%CITATION = ARXIV:0812.1172;%%.

\bibitem{Adams:2005ph}
\hrefCMSnoop {}{{STAR} Collaboration, ``{Distributions of charged hadrons
  associated with high transverse momentum particles in pp and Au+Au collisions
  at $\sqrt{s_{NN}}$ = 200 GeV}'',} \textit{ Phys. Rev. Lett.} \textbf{ 95}
  (2005) 152301,
  \href{http://dx.doi.org/10.1103/PhysRevLett.95.152301}{\doi{10.1103/PhysRevLett.95.152301}},
\href{http://www.arXiv.org/abs/nucl-ex/0501016}{\texttt{arXiv:nucl-ex/0501016}}.
%%CITATION = NUCL-EX/0501016;%%.

\bibitem{Abelev:2009af}
\hrefCMSnoop {}{{STAR} Collaboration, ``{Long range rapidity correlations and
  jet production in high energy nuclear collisions}'',} \textit{ Phys. Rev. C}
  \textbf{ 80} (2009) 064912,
  \href{http://dx.doi.org/10.1103/PhysRevC.80.064912}{\doi{10.1103/PhysRevC.80.064912}},
\href{http://www.arXiv.org/abs/0909.0191}{\texttt{arXiv:0909.0191}}.
%%CITATION = ARXIV:0909.0191;%%.

\bibitem{Alver:2009id}
\hrefCMSnoop {}{{PHOBOS} Collaboration, ``{High transverse momentum triggered
  correlations over a large pseudorapidity acceptance in Au+Au collisions at
  $\sqrt{s_{NN}}$ = 200 GeV}'',} \textit{ Phys. Rev. Lett.} \textbf{ 104}
  (2010) 062301,
  \href{http://dx.doi.org/10.1103/PhysRevLett.104.062301}{\doi{10.1103/PhysRevLett.104.062301}},
\href{http://www.arXiv.org/abs/0903.2811}{\texttt{arXiv:0903.2811}}.
%%CITATION = ARXIV:0903.2811;%%.

\bibitem{Chatrchyan:2011eka}
\hrefCMSnoop {}{{CMS Collaboration}, ``{Long-range and short-range dihadron
  angular correlations in central PbPb collisions at a nucleon-nucleon center
  of mass energy of 2.76 TeV}'',} \textit{ JHEP} \textbf{ 07} (2011) 076,
  \href{http://dx.doi.org/10.1007/JHEP07(2011)076}{\doi{10.1007/JHEP07(2011)076}},
\href{http://www.arXiv.org/abs/1105.2438}{\texttt{arXiv:1105.2438}}.
%%CITATION = ARXIV:1105.2438;%%.

\bibitem{ALICE:2011ab}
\hrefCMSnoop {}{{ALICE Collaboration}, ``Higher harmonic anisotropic flow
  measurements of charged particles in {Pb-Pb} collisions at
  $\sqrt{s_{NN}}$=2.76 {TeV}'',} \textit{ Phys. Rev. Lett.} \textbf{ 107}
  (2011) 032301,
  \href{http://dx.doi.org/10.1103/PhysRevLett.107.032301}{\doi{10.1103/PhysRevLett.107.032301}},
\href{http://www.arXiv.org/abs/1105.3865}{\texttt{arXiv:1105.3865}}.
%%CITATION = ARXIV:1105.3865;%%.

\bibitem{Aamodt:2011by}
\hrefCMSnoop {}{{ALICE Collaboration}, ``Harmonic decomposition of two-particle
  angular correlations in {Pb-Pb} collisions at $\sqrt{s_{NN}}=$ 2.76 {TeV}'',}
  \textit{ Phys. Lett. B} \textbf{ 708} (2012) 249,
  \href{http://dx.doi.org/10.1016/j.physletb.2012.01.060}{\doi{10.1016/j.physletb.2012.01.060}},
\href{http://www.arXiv.org/abs/1109.2501}{\texttt{arXiv:1109.2501}}.
%%CITATION = ARXIV:1109.2501;%%.

\bibitem{Chatrchyan:2012wg}
\hrefCMSnoop {}{{CMS Collaboration}, ``{Centrality dependence of dihadron
  correlations and azimuthal anisotropy harmonics in PbPb collisions at
  \sqrtsNN = 2.76 TeV}'',} \textit{ Eur. Phys. J. C} \textbf{ 72} (2012) 2012,
  \href{http://dx.doi.org/10.1140/epjc/s10052-012-2012-3}{\doi{10.1140/epjc/s10052-012-2012-3}},
\href{http://www.arXiv.org/abs/1201.3158}{\texttt{arXiv:1201.3158}}.
%%CITATION = ARXIV:1201.3158;%%.

\bibitem{Aamodt:2010pa}
\hrefCMSnoop {}{{ALICE Collaboration}, ``{Elliptic flow of charged particles in
  Pb-Pb collisions at 2.76 TeV}'',} \textit{ Phys. Rev. Lett.} \textbf{ 105}
  (2010) 252302,
  \href{http://dx.doi.org/10.1103/PhysRevLett.105.252302}{\doi{10.1103/PhysRevLett.105.252302}},
\href{http://www.arXiv.org/abs/1011.3914}{\texttt{arXiv:1011.3914}}.
%%CITATION = ARXIV:1011.3914;%%.

\bibitem{ATLAS:2012at}
\hrefCMSnoop {}{{ATLAS Collaboration}, ``{Measurement of the azimuthal
  anisotropy for charged particle production in $\sqrt{s_{NN}}$ = 2.76 TeV
  lead-lead collisions with the ATLAS detector}'',} \textit{ Phys. Rev. C}
  \textbf{ 86} (2012) 014907,
  \href{http://dx.doi.org/10.1103/PhysRevC.86.014907}{\doi{10.1103/PhysRevC.86.014907}},
\href{http://www.arXiv.org/abs/1203.3087}{\texttt{arXiv:1203.3087}}.
%%CITATION = ARXIV:1203.3087;%%.

\bibitem{Chatrchyan:2012ta}
\hrefCMSnoop {}{{CMS Collaboration}, ``{Measurement of the elliptic anisotropy
  of charged particles produced in PbPb collisions at $\sqrt{s_{NN}}$ = 2.76
  TeV}'',} \textit{ Phys. Rev. C} \textbf{ 87} (2013) 014902,
  \href{http://dx.doi.org/10.1103/PhysRevC.87.014902}{\doi{10.1103/PhysRevC.87.014902}},
\href{http://www.arXiv.org/abs/1204.1409}{\texttt{arXiv:1204.1409}}.
%%CITATION = ARXIV:1204.1409;%%.

\bibitem{Busza:2018rrf}
\hrefCMSnoop {}{W.~Busza, K.~Rajagopal, and W.~van~der Schee, ``{Heavy ion
  collisions: the big picture, and the big questions}'',} \textit{ Ann. Rev.
  Nucl. Part. Sci.} \textbf{ 68} (2018) 339,
  \href{http://dx.doi.org/10.1146/annurev-nucl-101917-020852}{\doi{10.1146/annurev-nucl-101917-020852}},
\href{http://www.arXiv.org/abs/1802.04801}{\texttt{arXiv:1802.04801}}.
%%CITATION = ARXIV:1802.04801;%%.

\bibitem{Alver:2010dn}
\hrefCMSnoop {}{B.~H. Alver, C.~Gombeaud, M.~Luzum, and J.-Y. Ollitrault,
  ``{Triangular flow in hydrodynamics and transport theory}'',} \textit{ Phys.
  Rev. C} \textbf{ 82} (2010) 034913,
  \href{http://dx.doi.org/10.1103/PhysRevC.82.034913}{\doi{10.1103/PhysRevC.82.034913}},
\href{http://www.arXiv.org/abs/1007.5469}{\texttt{arXiv:1007.5469}}.
%%CITATION = ARXIV:1007.5469;%%.

\bibitem{Schenke:2010rr}
\hrefCMSnoop {}{B.~Schenke, S.~Jeon, and C.~Gale, ``{Elliptic and triangular
  flow in event-by-event (3+1)D viscous hydrodynamics}'',} \textit{ Phys. Rev.
  Lett.} \textbf{ 106} (2011) 042301,
  \href{http://dx.doi.org/10.1103/PhysRevLett.106.042301}{\doi{10.1103/PhysRevLett.106.042301}},
\href{http://www.arXiv.org/abs/1009.3244}{\texttt{arXiv:1009.3244}}.
%%CITATION = ARXIV:1009.3244;%%.

\bibitem{Qiu:2011hf}
\hrefCMSnoop {}{Z.~Qiu, C.~Shen, and U.~Heinz, ``{Hydrodynamic elliptic and
  triangular flow in Pb-Pb collisions at $\sqrt{s_{NN}} $ = 2.76 ATeV}'',}
  \textit{ Phys. Lett. B} \textbf{ 707} (2012) 151,
  \href{http://dx.doi.org/10.1016/j.physletb.2011.12.041}{\doi{10.1016/j.physletb.2011.12.041}},
\href{http://www.arXiv.org/abs/1110.3033}{\texttt{arXiv:1110.3033}}.
%%CITATION = ARXIV:1110.3033;%%.

\bibitem{Alver:2010gr}
\hrefCMSnoop {}{B.~Alver and G.~Roland, ``{Collision geometry fluctuations and
  triangular flow in heavy-ion collisions}'',} \textit{ Phys. Rev. C} \textbf{
  81} (2010) 054905,
  \href{http://dx.doi.org/10.1103/PhysRevC.81.054905}{\doi{10.1103/PhysRevC.81.054905}},
  \href{http://www.arXiv.org/abs/1003.0194}{\texttt{arXiv:1003.0194}}.
[Erratum: \DOI{10.1103/PhysRevC.82.039903}].
%%CITATION = ARXIV:1003.0194;%%.

\bibitem{Miller:2007ri}
\hrefCMSnoop {}{M.~L. Miller, K.~Reygers, S.~J. Sanders, and P.~Steinberg,
  ``{Glauber modeling in high energy nuclear collisions}'',} \textit{ Ann. Rev.
  Nucl. Part. Sci.} \textbf{ 57} (2007) 205,
  \href{http://dx.doi.org/10.1146/annurev.nucl.57.090506.123020}{\doi{10.1146/annurev.nucl.57.090506.123020}},
\href{http://www.arXiv.org/abs/nucl-ex/0701025}{\texttt{arXiv:nucl-ex/0701025}}.
%%CITATION = NUCL-EX/0701025;%%.

\bibitem{Khachatryan:2010gv}
\hrefCMSnoop {}{{CMS Collaboration}, ``Observation of long-range near-side
  angular correlations in proton-proton collisions at the {LHC}'',} \textit{
  JHEP} \textbf{ 09} (2010) 091,
  \href{http://dx.doi.org/10.1007/JHEP09(2010)091}{\doi{10.1007/JHEP09(2010)091}},
\href{http://www.arXiv.org/abs/1009.4122}{\texttt{arXiv:1009.4122}}.
%%CITATION = ARXIV:1009.4122;%%.

\bibitem{Aad:2015gqa}
\hrefCMSnoop {}{{ATLAS Collaboration}, ``Observation of long-range elliptic
  azimuthal anisotropies in $\sqrt{s}$ = 13 and 2.76 {TeV} $pp$ collisions with
  the {ATLAS} detector'',} \textit{ Phys. Rev. Lett.} \textbf{ 116} (2016)
  172301,
  \href{http://dx.doi.org/10.1103/PhysRevLett.116.172301}{\doi{10.1103/PhysRevLett.116.172301}},
\href{http://www.arXiv.org/abs/1509.04776}{\texttt{arXiv:1509.04776}}.
%%CITATION = ARXIV:1509.04776;%%.

\bibitem{Khachatryan:2015lva}
\hrefCMSnoop {}{{CMS Collaboration}, ``{Measurement of long-range near-side
  two-particle angular correlations in pp collisions at $\sqrt{s}$ = 13
  TeV}'',} \textit{ Phys. Rev. Lett.} \textbf{ 116} (2016) 172302,
  \href{http://dx.doi.org/10.1103/PhysRevLett.116.172302}{\doi{10.1103/PhysRevLett.116.172302}},
\href{http://www.arXiv.org/abs/1510.03068}{\texttt{arXiv:1510.03068}}.
%%CITATION = ARXIV:1510.03068;%%.

\bibitem{CMS:2012qk}
\hrefCMSnoop {}{{CMS Collaboration}, ``{Observation of long-range near-side
  angular correlations in proton-lead collisions at the LHC}'',} \textit{ Phys.
  Lett. B} \textbf{ 718} (2013) 795,
  \href{http://dx.doi.org/10.1016/j.physletb.2012.11.025}{\doi{10.1016/j.physletb.2012.11.025}},
\href{http://www.arXiv.org/abs/1210.5482}{\texttt{arXiv:1210.5482}}.
%%CITATION = ARXIV:1210.5482;%%.

\bibitem{Abelev:2012ola}
\hrefCMSnoop {}{{ALICE Collaboration}, ``{Long-range angular correlations on
  the near and away side in $p$-Pb collisions at $\sqrt{s_{NN}}$ = 5.02
  TeV}'',} \textit{ Phys. Lett. B} \textbf{ 719} (2013) 29,
  \href{http://dx.doi.org/10.1016/j.physletb.2013.01.012}{\doi{10.1016/j.physletb.2013.01.012}},
\href{http://www.arXiv.org/abs/1212.2001}{\texttt{arXiv:1212.2001}}.
%%CITATION = ARXIV:1212.2001;%%.

\bibitem{Aad:2012gla}
\hrefCMSnoop {}{{ATLAS Collaboration}, ``Observation of associated near-side
  and away-side long-range correlations in $\sqrt{s_{NN}}$ = 5.02 {TeV}
  proton-lead collisions with the {ATLAS} detector'',} \textit{ Phys. Rev.
  Lett.} \textbf{ 110} (2013) 182302,
  \href{http://dx.doi.org/10.1103/PhysRevLett.110.182302}{\doi{10.1103/PhysRevLett.110.182302}},
\href{http://www.arXiv.org/abs/1212.5198}{\texttt{arXiv:1212.5198}}.
%%CITATION = ARXIV:1212.5198;%%.

\bibitem{Aaij:2015qcq}
\hrefCMSnoop {}{{LHCb Collaboration}, ``{Measurements of long-range near-side
  angular correlations in $\sqrt{s_{\text{NN}}}$ = 5 TeV proton-lead collisions
  in the forward region}'',} \textit{ Phys. Lett. B} \textbf{ 762} (2016) 473,
  \href{http://dx.doi.org/10.1016/j.physletb.2016.09.064}{\doi{10.1016/j.physletb.2016.09.064}},
\href{http://www.arXiv.org/abs/1512.00439}{\texttt{arXiv:1512.00439}}.
%%CITATION = ARXIV:1512.00439;%%.

\bibitem{Aidala:2017ajz}
\hrefCMSnoop {}{{PHENIX} Collaboration, ``{Measurements of multiparticle
  correlations in $d$$+$$\mathrm{Au}$ collisions at 200, 62.4, 39, and 19.6 GeV
  and $p$$+$$\mathrm{Au}$ collisions at 200 GeV and implications for collective
  behavior}'',} \textit{ Phys. Rev. Lett.} \textbf{ 120} (2018) 062302,
  \href{http://dx.doi.org/10.1103/PhysRevLett.120.062302}{\doi{10.1103/PhysRevLett.120.062302}},
\href{http://www.arXiv.org/abs/1707.06108}{\texttt{arXiv:1707.06108}}.
%%CITATION = ARXIV:1707.06108;%%.

\bibitem{Aidala:2016vgl}
\hrefCMSnoop {}{{PHENIX} Collaboration, ``{Measurement of long-range angular
  correlations and azimuthal anisotropies in high-multiplicity $p$$+$Au
  collisions at $\sqrt{s_{_{NN}}}=200$ GeV}'',} \textit{ Phys. Rev. C} \textbf{
  95} (2017) 034910,
  \href{http://dx.doi.org/10.1103/PhysRevC.95.034910}{\doi{10.1103/PhysRevC.95.034910}},
\href{http://www.arXiv.org/abs/1609.02894}{\texttt{arXiv:1609.02894}}.
%%CITATION = ARXIV:1609.02894;%%.

\bibitem{Adare:2015ctn}
\hrefCMSnoop {}{{PHENIX} Collaboration, ``{Measurements of elliptic and
  triangular flow in high-multiplicity $^{3}$He$+$Au collisions at
  $\sqrt{s_{_{NN}}}=200$ GeV}'',} \textit{ Phys. Rev. Lett.} \textbf{ 115}
  (2015) 142301,
  \href{http://dx.doi.org/10.1103/PhysRevLett.115.142301}{\doi{10.1103/PhysRevLett.115.142301}},
\href{http://www.arXiv.org/abs/1507.06273}{\texttt{arXiv:1507.06273}}.
%%CITATION = ARXIV:1507.06273;%%.

\bibitem{PHENIX:2018lia}
\hrefCMSnoop {}{{PHENIX} Collaboration, ``{Creation of quark-gluon plasma
  droplets with three distinct geometries}'',} \textit{ Nature Phys.} \textbf{
  15} (2019) 214,
  \href{http://dx.doi.org/10.1038/s41567-018-0360-0}{\doi{10.1038/s41567-018-0360-0}},
\href{http://www.arXiv.org/abs/1805.02973}{\texttt{arXiv:1805.02973}}.
%%CITATION = ARXIV:1805.02973;%%.

\bibitem{Nagle:2018nvi}
\hrefCMSnoop {}{J.~L. Nagle and W.~A. Zajc, ``Small system collectivity in
  relativistic hadronic and nuclear collisions'',} \textit{ Ann. Rev. Nucl.
  Part. Sci.} \textbf{ 68} (2018) 211,
  \href{http://dx.doi.org/10.1146/annurev-nucl-101916-123209}{\doi{10.1146/annurev-nucl-101916-123209}},
\href{http://www.arXiv.org/abs/1801.03477}{\texttt{arXiv:1801.03477}}.
%%CITATION = ARXIV:1801.03477;%%.

\bibitem{Schenke:2014zha}
\hrefCMSnoop {}{B.~Schenke and R.~Venugopalan, ``{Eccentric protons?
  Sensitivity of flow to system size and shape in p+p, p+Pb and Pb+Pb
  collisions}'',} \textit{ Phys. Rev. Lett.} \textbf{ 113} (2014) 102301,
  \href{http://dx.doi.org/10.1103/PhysRevLett.113.102301}{\doi{10.1103/PhysRevLett.113.102301}},
\href{http://www.arXiv.org/abs/1405.3605}{\texttt{arXiv:1405.3605}}.
%%CITATION = ARXIV:1405.3605;%%.

\bibitem{Bozek:2011if}
\hrefCMSnoop {}{P.~Bozek, ``{Collective flow in p-Pb and d-Pd collisions at TeV
  energies}'',} \textit{ Phys. Rev. C} \textbf{ 85} (2012) 014911,
  \href{http://dx.doi.org/10.1103/PhysRevC.85.014911}{\doi{10.1103/PhysRevC.85.014911}},
\href{http://www.arXiv.org/abs/1112.0915}{\texttt{arXiv:1112.0915}}.
%%CITATION = ARXIV:1112.0915;%%.

\bibitem{Bozek:2012gr}
\hrefCMSnoop {}{P.~Bozek and W.~Broniowski, ``{Correlations from hydrodynamic
  flow in p-Pb collisions}'',} \textit{ Phys. Lett. B} \textbf{ 718} (2013)
  1557,
  \href{http://dx.doi.org/10.1016/j.physletb.2012.12.051}{\doi{10.1016/j.physletb.2012.12.051}},
\href{http://www.arXiv.org/abs/1211.0845}{\texttt{arXiv:1211.0845}}.
%%CITATION = ARXIV:1211.0845;%%.

\bibitem{Dusling:2012wy}
\hrefCMSnoop {}{K.~Dusling and R.~Venugopalan, ``{Explanation of systematics of
  CMS p+Pb high multiplicity di-hadron data at $\sqrt{s_{\rm NN}}$ = 5.02
  TeV}'',} \textit{ Phys. Rev. D} \textbf{ 87} (2013) 054014,
  \href{http://dx.doi.org/10.1103/PhysRevD.87.054014}{\doi{10.1103/PhysRevD.87.054014}},
\href{http://www.arXiv.org/abs/1211.3701}{\texttt{arXiv:1211.3701}}.
%%CITATION = ARXIV:1211.3701;%%.

\bibitem{Dusling:2017dqg}
\hrefCMSnoop {}{K.~Dusling, M.~Mace, and R.~Venugopalan, ``{Multiparticle
  collectivity from initial state correlations in high energy proton-nucleus
  collisions}'',} \textit{ Phys. Rev. Lett.} \textbf{ 120} (2018) 042002,
  \href{http://dx.doi.org/10.1103/PhysRevLett.120.042002}{\doi{10.1103/PhysRevLett.120.042002}},
\href{http://www.arXiv.org/abs/1705.00745}{\texttt{arXiv:1705.00745}}.
%%CITATION = ARXIV:1705.00745;%%.

\bibitem{Dusling:2017aot}
\hrefCMSnoop {}{K.~Dusling, M.~Mace, and R.~Venugopalan, ``{Parton model
  description of multiparticle azimuthal correlations in $pA$ collisions}'',}
  \textit{ Phys. Rev. D} \textbf{ 97} (2018) 016014,
  \href{http://dx.doi.org/10.1103/PhysRevD.97.016014}{\doi{10.1103/PhysRevD.97.016014}},
\href{http://www.arXiv.org/abs/1706.06260}{\texttt{arXiv:1706.06260}}.
%%CITATION = ARXIV:1706.06260;%%.

\bibitem{Borghini:2000sa}
\hrefCMSnoop {}{N.~Borghini, P.~M. Dinh, and J.-Y. Ollitrault, ``{A new method
  for measuring azimuthal distributions in nucleus-nucleus collisions}'',}
  \textit{ Phys. Rev. C} \textbf{ 63} (2001) 054906,
  \href{http://dx.doi.org/10.1103/PhysRevC.63.054906}{\doi{10.1103/PhysRevC.63.054906}},
\href{http://www.arXiv.org/abs/nucl-th/0007063}{\texttt{arXiv:nucl-th/0007063}}.
%%CITATION = NUCL-TH/0007063;%%.

\bibitem{Khachatryan:2016txc}
\hrefCMSnoop {}{{CMS Collaboration}, ``{Evidence for collectivity in pp
  collisions at the LHC}'',} \textit{ Phys. Lett. B} \textbf{ 765} (2017) 193,
  \href{http://dx.doi.org/10.1016/j.physletb.2016.12.009}{\doi{10.1016/j.physletb.2016.12.009}},
\href{http://www.arXiv.org/abs/1606.06198}{\texttt{arXiv:1606.06198}}.
%%CITATION = ARXIV:1606.06198;%%.

\bibitem{Khachatryan:2015waa}
\hrefCMSnoop {}{{CMS Collaboration}, ``{Evidence for collective multiparticle
  correlations in p-Pb collisions}'',} \textit{ Phys. Rev. Lett.} \textbf{ 115}
  (2015) 012301,
  \href{http://dx.doi.org/10.1103/PhysRevLett.115.012301}{\doi{10.1103/PhysRevLett.115.012301}},
\href{http://www.arXiv.org/abs/1502.05382}{\texttt{arXiv:1502.05382}}.
%%CITATION = ARXIV:1502.05382;%%.

\bibitem{Chatrchyan:2013kba}
\hrefCMSnoop {}{{CMS Collaboration}, ``{Measurement of higher-order harmonic
  azimuthal anisotropy in PbPb collisions at $\sqrt{s_{NN}}$ = 2.76 TeV}'',}
  \textit{ Phys. Rev. C} \textbf{ 89} (2014) 044906,
  \href{http://dx.doi.org/10.1103/PhysRevC.89.044906}{\doi{10.1103/PhysRevC.89.044906}},
\href{http://www.arXiv.org/abs/1310.8651}{\texttt{arXiv:1310.8651}}.
%%CITATION = ARXIV:1310.8651;%%.

\bibitem{Chatrchyan:2013nka}
\hrefCMSnoop {}{{CMS Collaboration}, ``{Multiplicity and transverse momentum
  dependence of two- and four-particle correlations in pPb and PbPb
  collisions}'',} \textit{ Phys. Lett. B} \textbf{ 724} (2013) 213,
  \href{http://dx.doi.org/10.1016/j.physletb.2013.06.028}{\doi{10.1016/j.physletb.2013.06.028}},
\href{http://www.arXiv.org/abs/1305.0609}{\texttt{arXiv:1305.0609}}.
%%CITATION = ARXIV:1305.0609;%%.

\bibitem{Bilandzic:2010jr}
\hrefCMSnoop {}{A.~Bilandzic, R.~Snellings, and S.~Voloshin, ``{Flow analysis
  with cumulants: Direct calculations}'',} \textit{ Phys. Rev. C} \textbf{ 83}
  (2011) 044913,
  \href{http://dx.doi.org/10.1103/PhysRevC.83.044913}{\doi{10.1103/PhysRevC.83.044913}},
\href{http://www.arXiv.org/abs/1010.0233}{\texttt{arXiv:1010.0233}}.
%%CITATION = ARXIV:1010.0233;%%.

\bibitem{Yan:2013laa}
\hrefCMSnoop {}{L.~Yan and J.-Y. Ollitrault, ``{Universal fluctuation-driven
  eccentricities in proton-proton, proton-nucleus and nucleus-nucleus
  collisions}'',} \textit{ Phys. Rev. Lett.} \textbf{ 112} (2014) 082301,
  \href{http://dx.doi.org/10.1103/PhysRevLett.112.082301}{\doi{10.1103/PhysRevLett.112.082301}},
\href{http://www.arXiv.org/abs/1312.6555}{\texttt{arXiv:1312.6555}}.
%%CITATION = ARXIV:1312.6555;%%.

\bibitem{Aaboud:2017blb}
\hrefCMSnoop {}{{ATLAS Collaboration}, ``Measurement of long-range
  multiparticle azimuthal correlations with the subevent cumulant method in
  $pp$ and {$p + Pb$} collisions with the {ATLAS} detector at the {CERN Large
  Hadron Collider}'',} \textit{ Phys. Rev. C} \textbf{ 97} (2018), no.~2,
  024904,
  \href{http://dx.doi.org/10.1103/PhysRevC.97.024904}{\doi{10.1103/PhysRevC.97.024904}},
\href{http://www.arXiv.org/abs/1708.03559}{\texttt{arXiv:1708.03559}}.
%%CITATION = ARXIV:1708.03559;%%.

\bibitem{Heinz:2013th}
\hrefCMSnoop {}{U.~Heinz and R.~Snellings, ``{Collective flow and viscosity in
  relativistic heavy-ion collisions}'',} \textit{ Ann. Rev. Nucl. Part. Sci.}
  \textbf{ 63} (2013) 123,
  \href{http://dx.doi.org/10.1146/annurev-nucl-102212-170540}{\doi{10.1146/annurev-nucl-102212-170540}},
\href{http://www.arXiv.org/abs/1301.2826}{\texttt{arXiv:1301.2826}}.
%%CITATION = ARXIV:1301.2826;%%.

\bibitem{Giacalone:2017uqx}
\hrefCMSnoop {}{G.~Giacalone, J.~Noronha-Hostler, and J.-Y. Ollitrault,
  ``{Relative flow fluctuations as a probe of initial state fluctuations}'',}
  \textit{ Phys. Rev. C} \textbf{ 95} (2017) 054910,
  \href{http://dx.doi.org/10.1103/PhysRevC.95.054910}{\doi{10.1103/PhysRevC.95.054910}},
\href{http://www.arXiv.org/abs/1702.01730}{\texttt{arXiv:1702.01730}}.
%%CITATION = ARXIV:1702.01730;%%.

\bibitem{Ollitrault:2009ie}
\hrefCMSnoop {}{J.-Y. Ollitrault, A.~M. Poskanzer, and S.~A. Voloshin,
  ``{Effect of flow fluctuations and nonflow on elliptic flow methods}'',}
  \textit{ Phys. Rev. C} \textbf{ 80} (2009) 014904,
  \href{http://dx.doi.org/10.1103/PhysRevC.80.014904}{\doi{10.1103/PhysRevC.80.014904}},
\href{http://www.arXiv.org/abs/0904.2315}{\texttt{arXiv:0904.2315}}.
%%CITATION = ARXIV:0904.2315;%%.

\bibitem{JINST}
\hrefCMSnoop {}{{CMS Collaboration}, ``The {CMS} experiment at the {CERN}
  {LHC}'',} \textit{ JINST} \textbf{ 3} (2008) S08004,
\href{http://dx.doi.org/10.1088/1748-0221/3/08/S08004}{\doi{10.1088/1748-0221/3/08/S08004}}.
%%CITATION = JINST,3,S08004;%%.

\bibitem{GEANT4}
\hrefCMSnoop {}{{GEANT4} Collaboration, ``{\GEANTfour}---a simulation
  toolkit'',} \textit{ Nucl. Instrum. Meth. A} \textbf{ 506} (2003) 250,
\href{http://dx.doi.org/10.1016/S0168-9002(03)01368-8}{\doi{10.1016/S0168-9002(03)01368-8}}.
%%CITATION = NUIMA,A506,250;%%.

\bibitem{CMS-PAS-LUM-17-002}
\href {http://cds.cern.ch/record/2628652}{{CMS Collaboration}, ``{CMS
  luminosity measurement using 2016 proton-nucleus collisions at
  nucleon-nucleon center-of-mass energy of 8.16 TeV}'',} CMS Physics Analysis
  Summary CMS-PAS-LUM-17-002, 2018.

\bibitem{Sirunyan:2017uyl}
\hrefCMSnoop {}{{CMS Collaboration}, ``{Observation of correlated azimuthal
  anisotropy fourier harmonics in pp and p+Pb collisions at the LHC}'',}
  \textit{ Phys. Rev. Lett.} \textbf{ 120} (2018) 092301,
  \href{http://dx.doi.org/10.1103/PhysRevLett.120.092301}{\doi{10.1103/PhysRevLett.120.092301}},
\href{http://www.arXiv.org/abs/1709.09189}{\texttt{arXiv:1709.09189}}.
%%CITATION = ARXIV:1709.09189;%%.

\bibitem{Khachatryan:2016bia}
\hrefCMSnoop {}{{CMS Collaboration}, ``{The CMS trigger system}'',} \textit{
  JINST} \textbf{ 12} (2017) 01020,
  \href{http://dx.doi.org/10.1088/1748-0221/12/01/P01020}{\doi{10.1088/1748-0221/12/01/P01020}},
\href{http://www.arXiv.org/abs/1609.02366}{\texttt{arXiv:1609.02366}}.
%%CITATION = ARXIV:1609.02366;%%.

\bibitem{Chatrchyan:2014fea}
\hrefCMSnoop {}{{CMS Collaboration}, ``{Description and performance of track
  and primary-vertex reconstruction with the CMS tracker}'',} \textit{ JINST}
  \textbf{ 9} (2014) P10009,
  \href{http://dx.doi.org/10.1088/1748-0221/9/10/P10009}{\doi{10.1088/1748-0221/9/10/P10009}},
\href{http://www.arXiv.org/abs/1405.6569}{\texttt{arXiv:1405.6569}}.
%%CITATION = ARXIV:1405.6569;%%.

\bibitem{Bilandzic:2013kga}
A.~Bilandzic\hrefCMSnoop {}{ {et~al.}, ``{Generic framework for anisotropic
  flow analyses with multi-particle azimuthal correlations}'',} \textit{ Phys.
  Rev. C} \textbf{ 89} (2014) 064904,
  \href{http://dx.doi.org/10.1103/PhysRevC.89.064904}{\doi{10.1103/PhysRevC.89.064904}},
\href{http://www.arXiv.org/abs/1312.3572}{\texttt{arXiv:1312.3572}}.
%%CITATION = ARXIV:1312.3572;%%.

\bibitem{Gyulassy:1994ew}
\hrefCMSnoop {}{M.~Gyulassy and X.-N. Wang, ``{HIJING 1.0: A Monte Carlo
  program for parton and particle production in high-energy hadronic and
  nuclear collisions}'',} \textit{ Comput. Phys. Commun.} \textbf{ 83} (1994)
  307,
  \href{http://dx.doi.org/10.1016/0010-4655(94)90057-4}{\doi{10.1016/0010-4655(94)90057-4}},
\href{http://www.arXiv.org/abs/nucl-th/9502021}{\texttt{arXiv:nucl-th/9502021}}.
%%CITATION = NUCL-TH/9502021;%%.

\bibitem{Sirunyan:2017igb}
\hrefCMSnoop {}{{CMS Collaboration}, ``{Pseudorapidity and transverse momentum
  dependence of flow harmonics in pPb and PbPb collisions}'',} \textit{ Phys.
  Rev. C} \textbf{ 98} (2018) 044902,
  \href{http://dx.doi.org/10.1103/PhysRevC.98.044902}{\doi{10.1103/PhysRevC.98.044902}},
\href{http://www.arXiv.org/abs/1710.07864}{\texttt{arXiv:1710.07864}}.
%%CITATION = ARXIV:1710.07864;%%.

\bibitem{Kozlov:2014fqa}
I.~Kozlov\hrefCMSnoop {}{ {et~al.}, ``{Transverse momentum structure of pair
  correlations as a signature of collective behavior in small collision
  systems}'',} (2014).
\href{http://www.arXiv.org/abs/1405.3976}{\texttt{arXiv:1405.3976}}.
%%CITATION = ARXIV:1405.3976;%%.

\bibitem{Bernhard:2016tnd}
J.~E. Bernhard\hrefCMSnoop {}{ {et~al.}, ``{Applying Bayesian parameter
  estimation to relativistic heavy-ion collisions: simultaneous
  characterization of the initial state and quark-gluon plasma medium}'',}
  \textit{ Phys. Rev. C} \textbf{ 94} (2016) 024907,
  \href{http://dx.doi.org/10.1103/PhysRevC.94.024907}{\doi{10.1103/PhysRevC.94.024907}},
\href{http://www.arXiv.org/abs/1605.03954}{\texttt{arXiv:1605.03954}}.
%%CITATION = ARXIV:1605.03954;%%.

\bibitem{Khachatryan:2015oea}
\hrefCMSnoop {}{{CMS Collaboration}, ``{Evidence for transverse momentum and
  pseudorapidity dependent event plane fluctuations in PbPb and pPb
  collisions}'',} \textit{ Phys. Rev. C} \textbf{ 92} (2015) 034911,
  \href{http://dx.doi.org/10.1103/PhysRevC.92.034911}{\doi{10.1103/PhysRevC.92.034911}},
\href{http://www.arXiv.org/abs/1503.01692}{\texttt{arXiv:1503.01692}}.
%%CITATION = ARXIV:1503.01692;%%.

\end{thebibliography}\endgroup

\cleardoublepage \appendix\section{The CMS Collaboration \label{app:collab}}\begin{sloppypar}\hyphenpenalty=5000\widowpenalty=500\clubpenalty=5000\vskip\cmsinstskip
\textbf{Yerevan Physics Institute, Yerevan, Armenia}\\*[0pt]
A.M.~Sirunyan, A.~Tumasyan
\vskip\cmsinstskip
\textbf{Institut f\"{u}r Hochenergiephysik, Wien, Austria}\\*[0pt]
W.~Adam, F.~Ambrogi, E.~Asilar, T.~Bergauer, J.~Brandstetter, M.~Dragicevic, J.~Er\"{o}, A.~Escalante~Del~Valle, M.~Flechl, R.~Fr\"{u}hwirth\cmsAuthorMark{1}, V.M.~Ghete, J.~Hrubec, M.~Jeitler\cmsAuthorMark{1}, N.~Krammer, I.~Kr\"{a}tschmer, D.~Liko, T.~Madlener, I.~Mikulec, N.~Rad, H.~Rohringer, J.~Schieck\cmsAuthorMark{1}, R.~Sch\"{o}fbeck, M.~Spanring, D.~Spitzbart, W.~Waltenberger, J.~Wittmann, C.-E.~Wulz\cmsAuthorMark{1}, M.~Zarucki
% ...
\vskip\cmsinstskip
% ...
1:  Also at Vienna University of Technology, Vienna, Austria\\
% ...
\end{sloppypar}
\end{document}